\documentstyle[times,pramana,epsf,floats,psfig,epsfig,twoside,amssymb,
rotating,colordvi,helvet,color,subfigure,axodraw]{ias}

\def\gsim{\buildrel{\scriptscriptstyle >}\over{\scriptscriptstyle\sim}}

\def\s#1{{\small#1}}

\def\HW{\s{HERWIG}}

\def\IS{\s{ISAJET}}

\def\SY{\s{SUSY}}

\def\B0bar{\overline{B^0}}

\begin{document}
\mark{{Improving the discovery potential of charged Higgs bosons ...}{Stefano Moretti}}

\title{Improving the discovery potential of charged Higgs bosons\\
at the Tevatron and Large Hadron Collider}

\author{Stefano Moretti}
\address{CERN Theory Division, CH-1211 Geneva 23, Switzerland
and\\
Institute for Particle Physics Phenomenology,
University of Durham, Durham DH1 3LE, UK}
\abstract{
We outline several improvements to the experimental analyses carried out 
at Tevatron (Run 2) or simulated in view of the Large Hadron Collider
(LHC) that could increase the scope of CDF/D0 and ATLAS/CMS in 
detecting charged Higgs bosons.}

\maketitle

\vskip-8.5cm

\noindent
\hskip9.65cm{CERN-TH/2002-076}

\noindent
\hskip9.65cm{IPPP/02/23}

\noindent
\hskip9.65cm{DCPT/02/46}

\noindent
\hskip9.65cm{April 2002}

\vskip+8.5cm

\section{Introduction}

The detection of charged Higgs bosons ($H^\pm$) at Tevatron or the LHC would
unequivocally imply the existence of physics beyond the Standard Model (SM),
since spin-less charged scalar states do not belong to its particle spectrum.
Singly charged Higgs bosons appear in any Two-Higgs Doublet Model (2HDM),
including a Type-II  in presence of minimal Supersymmetry (\SY), namely, 
the Minimal Supersymmetric Standard Model (MSSM). In the latter scenario,
these particles may be a unique probe of the `decoupling limit', wherein the 
lightest scalar Higgs boson of the MSSM, $h$, is completely
degenerate with the SM Higgs boson (i.e., same mass, couplings and physics
properties in the interaction with ordinary matter), the other four 
Higgs states of the model, $H$ (the heaviest scalar one),
$A$ (the pseudoscalar one) and the two charged ones, being much heavier,
likewise for the new \SY\ particles (squarks, sleptons and gauginos).
A valuable introduction to charged Higgs boson physics at hadron colliders
can be found in \cite{LesHouches}.

\begin{figure}[!t]
  \begin{center}
\vspace{-1.0cm}
    \parbox{2.in}{\epsfxsize=\hsize\epsffile{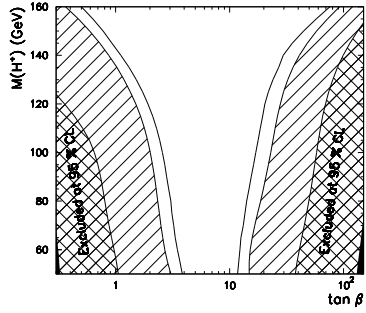}}
  \end{center}
\vspace{-0.25cm}
\caption{\small The D0/CDF combined 95\% CL exclusion boundaries in the [$M_{H^\pm},\tan\beta$] plane 
         for several values of the integrated luminosity:
         0.1 fb$^{-1}$ (at $\sqrt{s} = 1.8$ TeV, cross-hatched), 
         2.0 fb$^{-1}$ (at $\sqrt{s} = 2.0$ TeV, single-hatched) and 
         10 fb$^{-1}$ (at $\sqrt{s} = 2.0$ TeV, hollow).}
  \label{fig:Tevatron}
\end{figure}

\section{The top threshold region at Tevatron and the LHC}

\begin{figure}[!h]
\vspace{-0.25cm}
\begin{center}
\epsfig{file=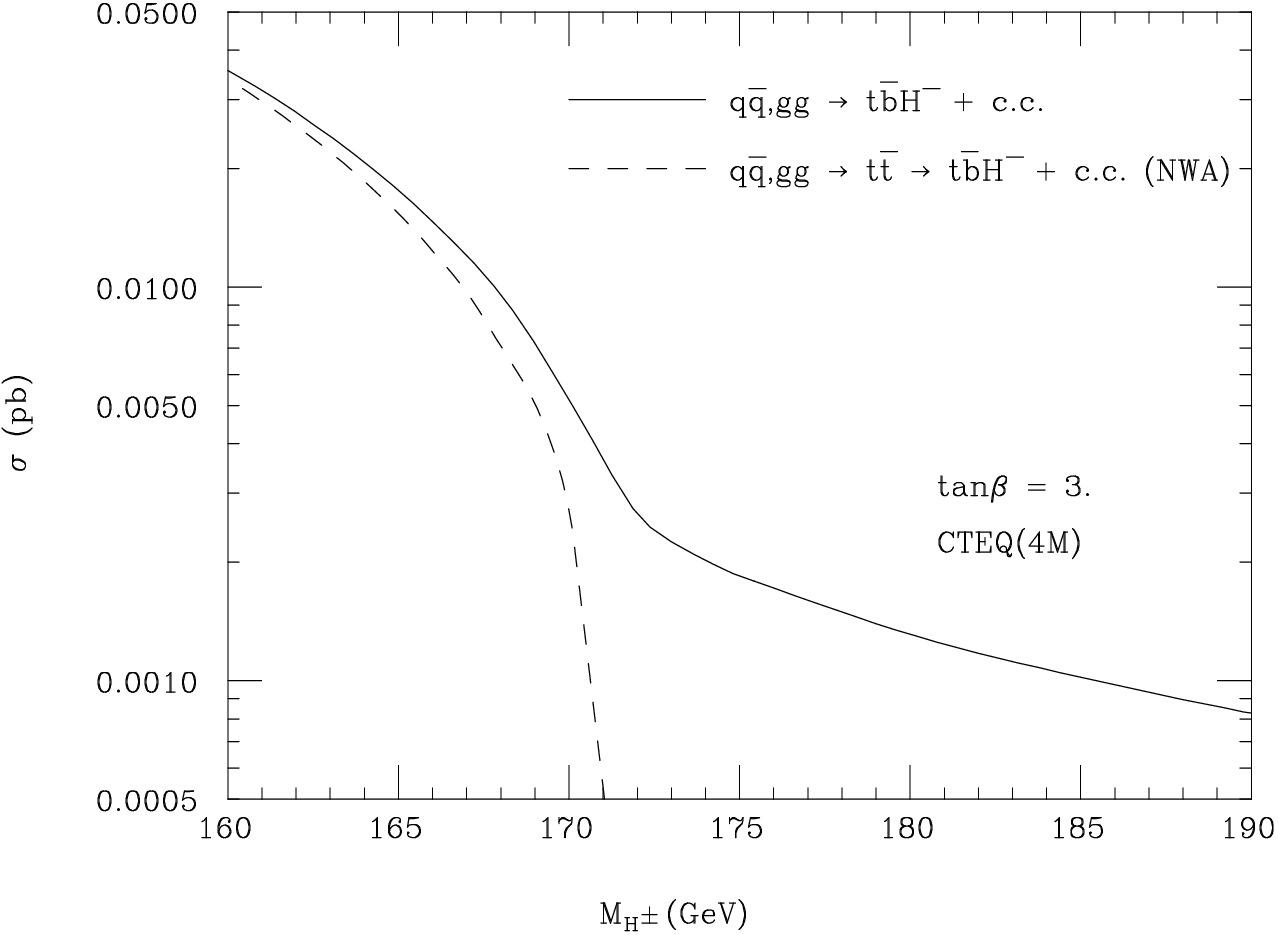,angle=90,height=4.5cm}
\epsfig{file=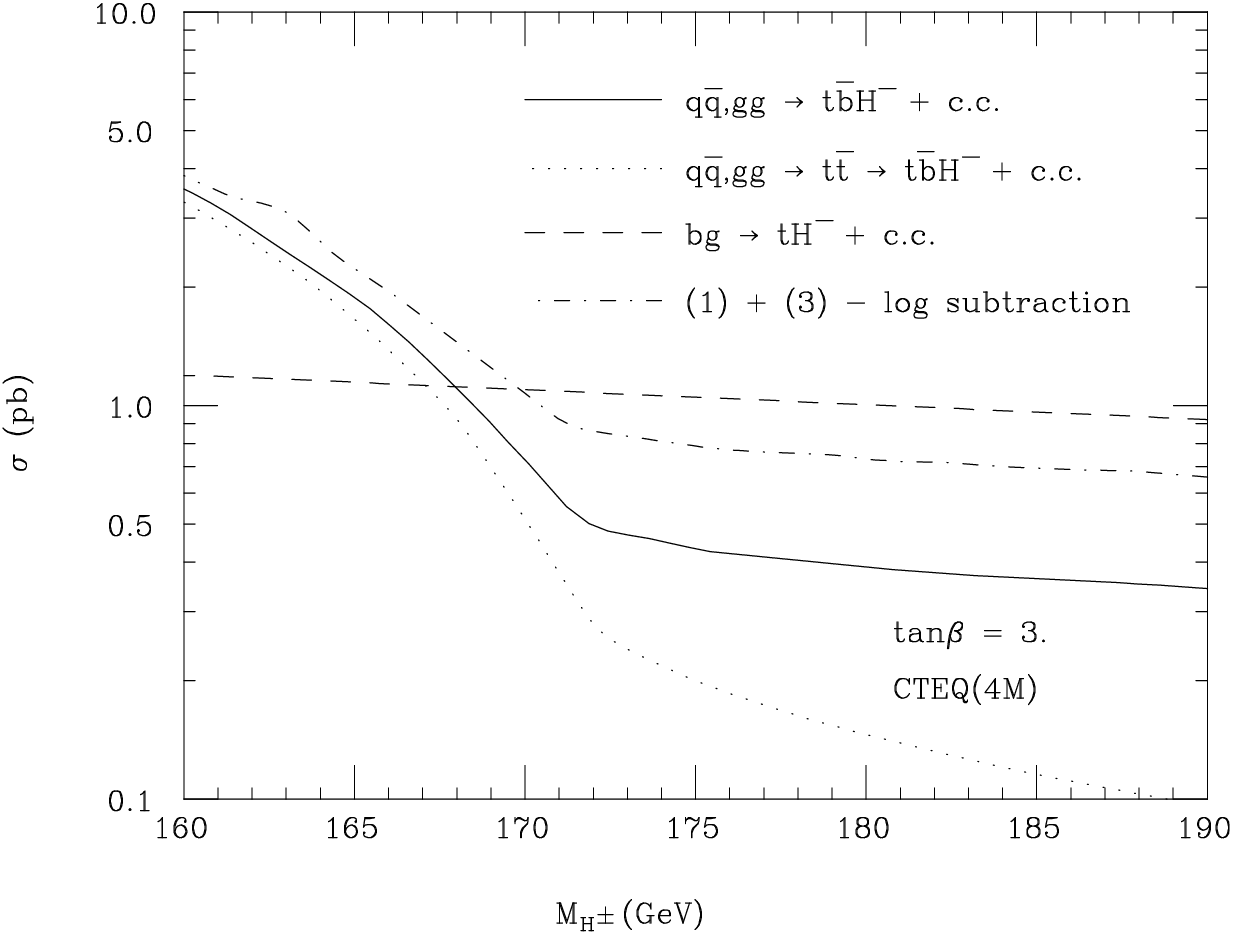,angle=90,height=4.5cm}
\caption{\small Cross section for $gg,q\bar q\to t\bar b H^-$ and
$gg,q\bar q \to t\bar t \to t\bar b H^-$ in NWA, at 
the Tevatron with $\sqrt s=2$ TeV (left plot). 
Cross section for $gg,q\bar q\to t\bar b H^-$,
$gg,q\bar q \to t\bar t \to t\bar b H^-$ with finite top quark width, 
$bg\to tH^-$ and
the combination of the first and the last, at the LHC with $\sqrt s=14$ TeV
(right plot). Rates are
function of $M_{H^\pm}$ for a representative value of $\tan\beta$.}
\label{fig:threshold}
\vspace{-0.25cm}
\end{center}
\end{figure}

The Run 2 discovery potential of $H^\pm$ bosons in a general Type-II 2HDM  is
visualised in Fig.~\ref{fig:Tevatron} (from Sect. II.G of Ref.~\cite{Run2}).
However, notice that the discovery reaches presented there ought to be 
considered as `conservative'. The reason being that they have been
assessed by running Monte Carlo (MC) simulations of $H^\pm$ production and
decay channels that may severely underestimate the actual scope
of charged Higgs boson searches. In fact,  those estimates were made by
assuming as  main production mode of $H^\pm$ scalars the decay of top
(anti)quarks produced via QCD in the annihilation
of gluon-gluon and quark-antiquark pairs (hence -- by definition -- 
the attainable Higgs mass is strictly confined to the region $M_{H^\pm}<m_t=175$ GeV).
This is not surprising, since standard MC programs, such as
{\small PYTHIA}, \HW\ and \IS\ \cite{PYTHIA,HERWIG,ISAJET}, have historically
accounted for this process through the usual procedure of factorising
the production mode, $gg,q\bar q\to t\bar t$, times the
decay one, $\bar t\to \bar b H^-$, in the so-called Narrow Width
Approximation (NWA) \cite{mono}. This description
fails to correctly account for the production 
phenomenology of charged Higgs bosons when their mass approaches or indeed
exceeds that of the top-quark (i.e., falls in the `threshold region',
$M_{H^\pm}\gsim m_t$) . This is evident from the left plot in 
Fig.~\ref{fig:threshold}. As remarked in
Ref.~\cite{mono}, the use of the $2\to 3$ hard scattering
process $gg,q\bar q\to t\bar b H^-$  \cite{tbH}, in place of the
`factorisation' procedure in NWA, is mandatory in the threshold
region, as the former correctly keeps into account 
both the effect of the finite width of the top quark
and the presence of other $H^\pm$ production mechanisms,
such Higgs-strahlung and $b\bar t \to H^-$ fusion (and relative interferences). 
The differences seen between the two descriptions in the left plot of 
Fig.~\ref{fig:threshold} are independent of $\tan\beta$ and 
also survive in, e.g., $p_T$ and $\eta$ spectra \cite{mono}.

If one then
looks at the most promising (and cleanest) charged Higgs boson decay 
channel, i.e., $H^\pm\to\tau^\pm\nu_\tau$~\cite{BRs}, while reconstructing the 
accompanying top quark hadronically, the prospects of $H^\pm$ detection 
improve significantly with increasing $M_{H^\pm}$ values. By following
the selection procedure outlined in Ref.~\cite{mono}, one can establish
at the Tevatron the presence of an $H^\pm$ signal 
over the dominant (irreducible) background 
(that is, $gg,q\bar q \to t\bar b W^-$ + c.c. events, yielding the same 
final state as the signal) up to masses of order $m_t$, hence in excess of 10
GeV or so with respect to 
 the values in Fig.~\ref{fig:Tevatron}, for the same choice of
$\tan\beta$: see Tab.~1 of \cite{mono}.  
The situation can be improved even further by taking advantage of 
$\tau$-polarisation effects, as explained in \cite{DP}. For example, 
by requiring that 80\% of the $\tau$-jet (transverse) energy is carried away 
by the $\pi^\pm$'s in one-prong decays, one can
reduce the background by a factor of 5, while costing to the signal 
only a more modest 50\% reduction (for any $M_{H^\pm}$ value
 between 160 GeV and $m_t$). 

\begin{figure}[!h]
\vspace{-0.25cm}
\begin{center}
\epsfig{file=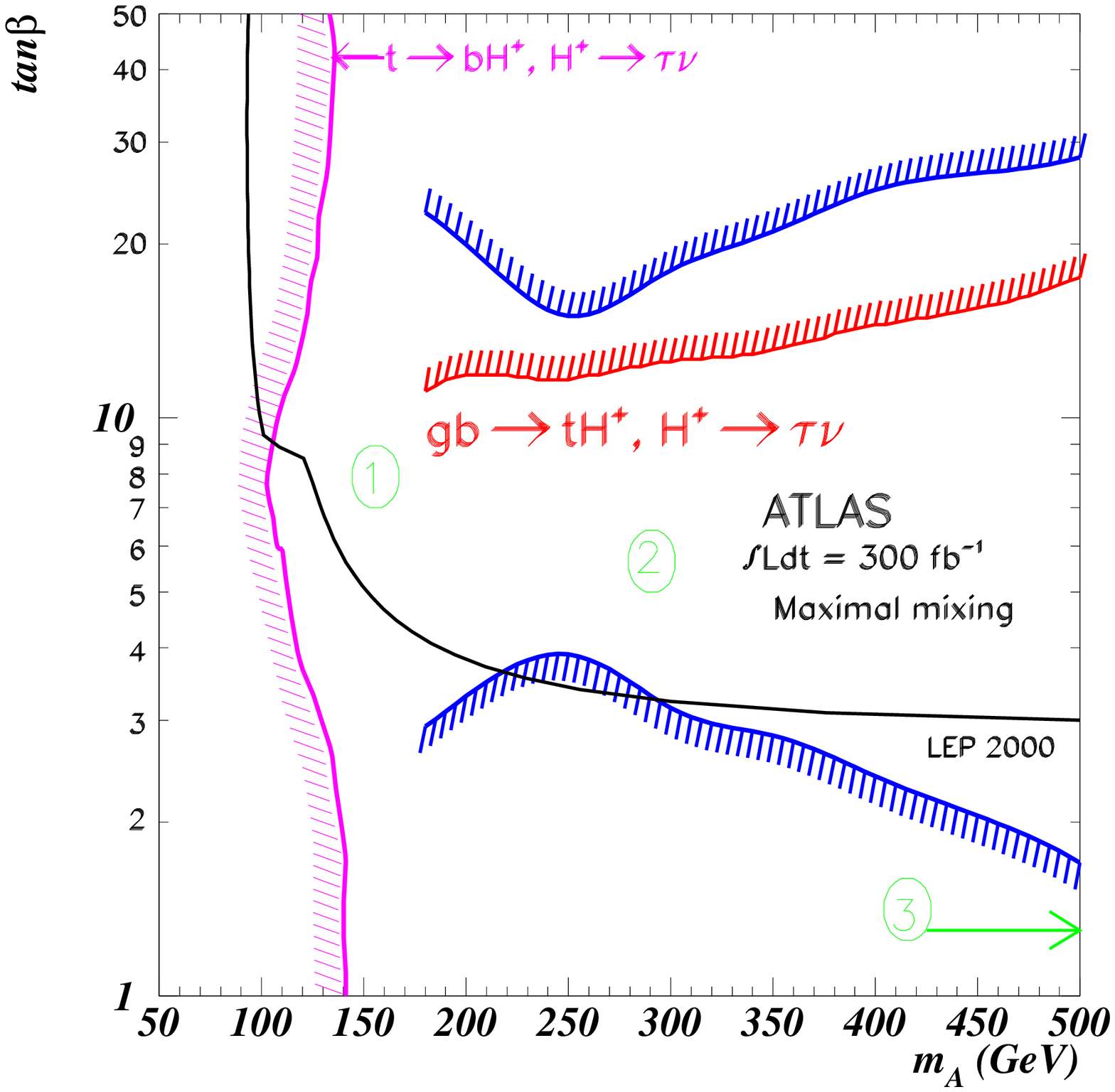,width=6.5cm,height=5.0cm}
\epsfig{file=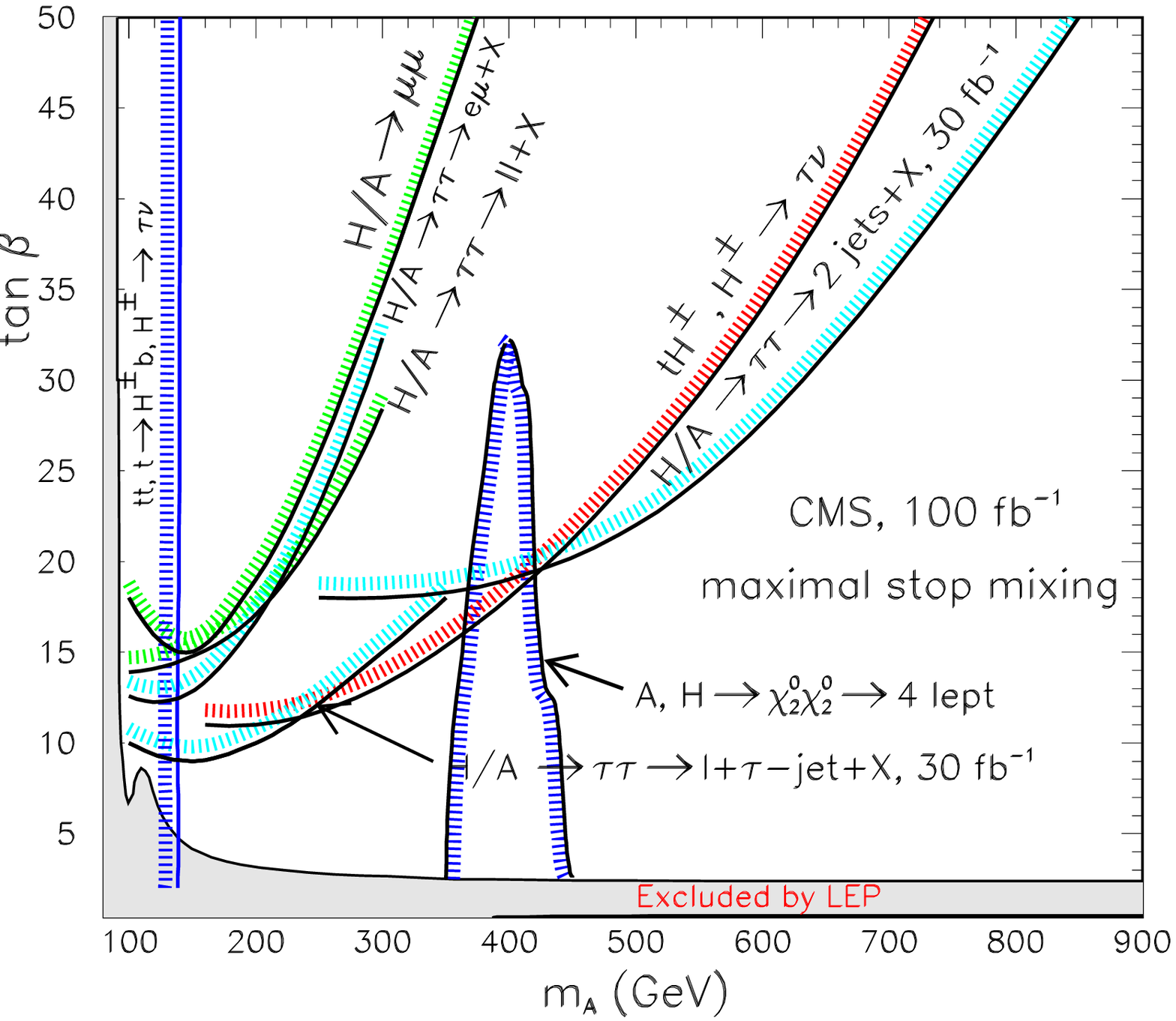,width=6.0cm,height=5.0cm}
\caption{\small The ATLAS 5-$\sigma$ discovery contours of 
2HDM charged Higgs 
bosons for 
300~fb$^{-1}$ of luminosity, only including the reach of SM decay modes  (left plot).
The CMS 5-$\sigma$ discovery contours of MSSM Higgs bosons for 100~fb$^{-1}$ of 
luminosity, also including the reach of $H,A \to \chi^0_2 \chi^0_2 \to 4l^{\pm}$ decays, assuming $M_1$ = 90~GeV,
$M_2$ = 180~GeV, $\mu$ = 500~GeV, $M_{\tilde{\ell}}$ = 250~GeV, $M_{\tilde{q},\tilde{g}}$ = 1000~GeV 
(right plot). }
\label{fig:LHC}
{\vskip-6.0cm\hskip-6.0cm{\tiny{\Blue{$gb\to tH^+$, $H^+\to t\bar b$}}}}
\vskip+5.0cm
\end{center}
\end{figure}

The problematic just illustrated for the case of the Tevatron
is very similar at the LHC, if anything more complicated.
In fact, at the CERN hadron collider, the above $2\to3$ reaction is
dominated by the $gg$-initiated subprocesses, rather than by 
$q\bar q$-annihilation, 
as is the case at the Tevatron. This means that a potential 
problem of double counting arises in the simulation of $t\bar b H^-$ + c.c. 
events at the LHC, if one considers that Higgs-strahlung can also be 
emulated through
the $2\to2$ process $bg\to t H^-$ + c.c., 
as was done in assessing the ATLAS discovery reaches 
in the $H^+\to t\bar b$ and $H^+\to \tau^+\nu_\tau$ 
channels \cite{KeteviReview}. The difference between the
two approaches is well understood, and a prescription exists for
combining the two, through the subtraction of a common logarithmic
term: see Refs.~\cite{subtraction,mor-roy,Jaume}. The right
plot in Fig.~\ref{fig:threshold} summarises all the discussed issues in the 
context of the LHC.
The  mentioned $2\to3$ description of the $H^\pm$ 
production dynamics and the spin correlations in $\tau$-decays are
now both available in version 6.4 of 
the \HW\  event generator (the latter also 
through an interface to {\small TAUOLA} \cite{TAUOLA}), so that detailed
simulations of $H^\pm$ signatures
at both the Tevatron and CERN hadron colliders are now possible
for the threshold region, including fragmentation/hadronisation and detector
effects. Its adoption will ultimately allow to `naturally' connect 
the discovery contours below and above the top threshold in the left plot of 
Fig.~\ref{fig:LHC}:
the uncovered area at $M_{H^\pm}\sim m_t$ (point 1.) is in fact
an artifact of the simulations adopted in ATLAS (the same occurs
in CMS: see right plot of Fig.~\ref{fig:LHC}).

\section{The intermediate $\tan\beta$ region at the LHC}

The second uncovered region at the LHC in the [$M_{H^\pm},\tan\beta$] plane
(see point 2. in the left plot of Fig.~\ref{fig:LHC}) is
precisely where the MSSM decoupling limit
onsets. A possible means of accessing this area of the
parameter space is to exploit \SY\ decays of charged Higgs bosons \cite{Mike},
similarly to what already done in CMS in the neutral sector (see right plot
of Fig.~\ref{fig:LHC}) \cite{neutralSUSY}. 
(For the impact of SUSY virtual effects see \cite{Jaume}.)
In particular, Ref.
\cite{Mike} showed
that intermediate values of $\tan\beta$ between $3$ and
$10$ could be in part accessible via $H^{\pm} \rightarrow
\widetilde{\chi}_1^{\pm} \widetilde{\chi}_{\{2,3\}}^0$
modes, resulting in three lepton final states (where leptons mean
electrons or muons), a hadronically reconstructed top quark
(from $gg \rightarrow \bar{b}tH^-$, $gb \rightarrow tH^-$
and their c.c. production processes) plus substantial
missing transverse momentum (from neutralino and chargino decays
to the stable lightest neutralino, $\widetilde{\chi}_{1}^0$, i.e., the lightest
supersymmetric particle or LSP).

These signals have preliminarily been looked at in the context of 
the 2001 Les Houches workshop (second paper of \cite{LesHouches}),
in presence of a full (CMS) detector simulation
(\HW\ 6.3 \cite{SUSYWIG} was used to generate all hard
processes). The results are 
rather promising, 
showing that all SM backgrounds can be completely
removed, leaving only MSSM processes as irreducible backgrounds in the
$3\ell+p^{\mathrm{miss}}_T+t$ channel ($\ell=e,\mu$).
Five MSSM points were considered, all in the intermediate $\tan\beta$ 
region: see top of
Tab.~\ref{tab:SUSY} (Here, $M_{\scriptscriptstyle 1} = 
\frac{1}{2}M_{\scriptscriptstyle 2}$ is
assumed).

\begin{table}[!h]
    \begin{center}
\vspace{-0.25cm}
    \caption{\small Top: Simulated MSSM parameter points (all masses in GeV).
The event number is the parton-level result for the production rates times
the Branching Ratios (BRs) for
$H^\pm \to \chi_1^\pm\chi^0_{\{2,3\}} \to 3\ell 
~p^{\mathrm{miss}}_T$ and $t\to bjj$ (where $j$
represent a non-$b$-jet).
Bottom: Number of events after cuts. All rates are given at a luminosity of $100\, \hbox{fb}^{-1}$.}
     {\begin{tabular}{lccccccc} \hline\hline
        Point  & $\tan\beta$ & $m_{H^{\pm}}$ & $\mu$ 
               & $M_{\scriptscriptstyle 2}$ 
               & $m_{\tilde{\ell}_{\scriptscriptstyle R}}$
               & $m_{\tilde{\ell}_{\scriptscriptstyle L}}$
               & events\\ \hline 
A &  $8$ & $250$ & $-115$ & $200$ & $120$ & $170$ & 1243\\ 
B & $10$ & $250$ & $-115$ & $200$ & $120$ & $170$ & 1521\\
C & $10$ & $300$ & $-115$ & $200$ & $120$ & $170$ & 1245\\
D & $10$ & $250$ & $+130$ & $210$ & $125$ & $175$ & 1288\\
E & $10$ & $300$ & $+130$ & $210$ & $125$ & $175$ & 1183\\
\hline\hline
      \end{tabular}}
    \end{center}
\vspace{-0.5cm}
    \begin{center}
     \begin{tabular}{lcccccc} \hline\hline
        Process          & 3$\ell$ events & $Z^0$-veto & 3,4 jets & $m_{jjj}\sim m_t$ & $M_{jj}\sim m_W^\dagger$  & others\\ \hline

    $t\bar{t}$             &    847      &  622  & 90 &   30      & 0    & 0                  \\
    $t\bar{t}Z^0$          &    244      &   34  & 13 &   5       & 0    & 0                  \\
    $t\bar{t}\gamma^*$     &    18       &   18  & 10 &   3       & 1    & 0                  \\
    $t\bar{t}h$            &    66       &   52  & 33 &   9       & 3    & 1                  \\
    $\tilde{\ell}\tilde{\ell}$ &    5007 &  4430 & 475& 112       & 2    & 0                    \\
    $\widetilde{\chi} \widetilde{\chi}$       &    8674      &  7047 & 1203 & 365 &19 &  3       \\
    $\tilde{q}$, $\tilde{g}$  & 37955    & 29484 &3507& 487  &100  & 0              \\    
    \hline
    $t H^+$ (point A)      &     251     &   241 & 80&   23     & 6 &   5            \\
    $t H^+$ (point B)      &     321     &   298 &118&   42     &13 &   9            \\
    $t H^+$ (point C)      &     279     &   258 &100&   36     &11 &   7            \\
    $t H^+$ (point D)      &     339     &   323 &121&   48     &13 &   9            \\
    $t H^+$ (point E)      &     291     &   278 &114&   40     &10 &   5            \\ \hline\hline
\end{tabular}
\label{tab:SUSY}

{\small
$^\dagger$Includes $b$-tagging efficiency for the third jet
 \phantom{aaaaaaaaaaaaaaaaaaaaaaaaaaaaaaaaaaaaaaaaaaaaaaaaa} }
\end{center}
\end{table}

(Other MSSM parameters were: 
$m_{\tilde{g}} = 700\, \hbox{GeV}$, $m_{\tilde{q}} = 1000\, \hbox{GeV}$,
$m_{\tilde{b}_{\scriptscriptstyle R}} = 800\, \hbox{GeV}$,
$m_{\tilde{t}_{\scriptscriptstyle L}} = 600\, \hbox{GeV}$,
$m_{\tilde{t}_{\scriptscriptstyle R}} = 500\, \hbox{GeV}$
and $A_t = 500\,  \hbox{GeV}$.
Notice that rather large gluino and squark masses are 
chosen to preclude 
charged Higgs boson production from MSSM cascade decays
\cite{asesh}, thus leaving the `direct' production modes discussed so far
as the only numerical relevant contributors at the LHC
\cite{hierarchy}.)

Following the selection criteria outlined in Sect. G of
the second paper in \cite{LesHouches}, one obtain the rates
reported at the bottom of
Tab.~\ref{tab:SUSY} Despite the limited $X t H^-$ (and the c.c.,
after the subtraction of the common term) production rate
precludes exploration for mass values larger than 
 $M_{H^\pm}~\sim~300\, \hbox{GeV}$, a signal could well be observed
above the background, provided that:
(i) $\mu$ and $M_{\scriptscriptstyle 2}$ are not much above the current
LEP restrictions from gaugino searches;
(ii) sleptons are sufficiently light (to enhance the 
$\widetilde{\chi}_{\{2,3\}}^0\to \widetilde{\chi}_{1}^0 \ell^+\ell^-$ 
decay rates). This is nonetheless a phenomenologically interesting
parameter configuration as it  will be promptly accessible at the LHC.
More simulations are however still needed to asses the real potential of SUSY
decays of charged Higgs bosons, without reducing the scope of
the SM decay modes, whose BRs can be suppressed by the opening of the new
channels.

\section{The heavy mass region at the LHC}

Point 3. in the left side of Fig.~\ref{fig:LHC} refers to
the possibility of increasing the $H^\pm$ discovery potential
of ATLAS and CMS in the $H^+\to t\bar b$ decay mode
to charged Higgs masses much heavier than those considered so far.
Following \cite{mor-roy,4b}, the key is to exploit kinematical
cuts on the $b$-quarks appearing in
\begin{equation}\label{proc}
gg,q\bar q\to b\bar t H^+\to b\bar b t\bar t \to n b jj \ell^\pm p_T^{\mathrm{miss}},
\end{equation}
where $n=3$ or 4, with $b$-quarks being tagged. (Notice that if $n=3$,
the usual subtraction procedure has to be implemented, after accounting
for the contribution from $g\bar b\to \bar t H^+$ and c.c. too, whereas $n=4$
implies that the `spectator' $b$-quark in the $2\to3$ mode has to
enter the detector region.)

\begin{figure}[!t]
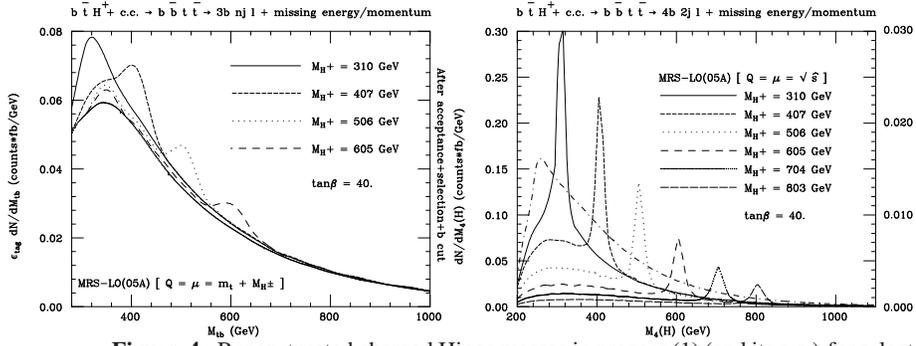

\vspace*{-12pt}
\begin{center}
\epsfig{file=newheavy.ps,angle=90,height=4.5cm}
\epsfig{file=reso4.ps,angle=90,height=4.5cm}
\caption{\small Reconstructed charged
Higgs masses in process (\ref{proc}) (and its c.c.) for
selected values of $M_{H^\pm}$ and  $\tan\beta=40$, after all cuts. 
Left plot is for the sum of signal and background, assuming 3 $b$-tags.
Right plot is for
signal and background (dot-dashed) separately, assuming 4 $b$-tags
(here, the right scale is obtained after multiplying by $\epsilon_b^4$).}
\label{fig:reso}
\vspace{-1.truecm}
\end{center}
\end{figure}

In both cases, an efficient $b$-tagging was assumed,
in order to get rid of QCD backgrounds in light-quark- and gluon-jets. 
Whereas this is possible in the case of 3 $b$-tags already for
a single $b$-tagging efficiency of $\epsilon_b\approx0.4$ 
for any $p_T(b)>30$ GeV, in the case of 4 $b$-tags the severe
suppression induced onto the acceptance rates for process
(\ref{proc}) by the requirement of detecting the spectator
$b$-quark imposes $\epsilon_b\approx0.56$ for any 
$p_T(b)>20$ GeV. If these performances can be achieved by 
the ATLAS and CMS detectors, then charged Higgs resonances can be extracted
from the backgrounds ($t\bar tj$, $t\bar t b$ and $t\bar t b\bar b$) 
with large statistical significances up to 600--800 GeV or so,
after simple kinematical cuts are applied on
the $b$-quarks not generated in the two (anti)top-quark decays.
Namely, by requiring either (i) $p_T(b_3)> 30$ GeV in the case
$n=3$ or (ii) $M_{b_3b_4}>120~{\mathrm{GeV}}$, $\cos\theta_{b_3b_4}<0.75$
and $E_{b_3}>120~{\mathrm{GeV}}$ in the case $n=4$ (where the subscripts
identify the $b$-quarks in terms of their decreasing energy:
i.e., $E_{b_3}>E_{b_4}$), one obtains the
encouraging parton-level results displayed in Fig.~\ref{fig:reso}
\cite{mor-roy,4b}.
Here, the normalisation is to the total cross section of (\ref{proc})
times the number of
possible `$2~b\, +\, 2$~jet mass' combinations: 
two in the left plot and and four in the right one.
These findings still await confirmation through more realistic experimental 
analyses,
but their potential in the high $\tan\beta$ region is clearly evident.

\end{document}